\documentclass[twocolumn]{aastex631}

\usepackage{amsmath}
\usepackage{graphicx}
\usepackage{textcomp}
\usepackage{gensymb}
\usepackage{natbib}
\usepackage{longtable}
\usepackage{multirow}
\usepackage{float}
\usepackage{enumitem}
\usepackage{anyfontsize}

\begin{document}
\def\teff{$T_{\rm eff}$}
\def\cs{$\chi^{2}$}
\def\rsun{$R_{\odot}$}
\def\msun{$M_{\odot}$}
\def\rstar{$R_{\star}$}
\def\rearth{$R_{\earth}$}
\def\av{$A_{V}$}
\def\emcee{\texttt{emcee}}
\def\kepler{\textit{Kepler}}

\graphicspath{{figures/}}
\shorttitle{Gaia Photometric Bias}
\shortauthors{Sullivan et al.}

\title{Quantifying the Contamination From Nearby Stellar Companions in Gaia DR3 Photometry}

\author[0000-0001-6873-8501]{Kendall Sullivan}
\affil{Department of Astronomy and Astrophysics, University of California Santa Cruz, Santa Cruz, CA 95064, USA}

\author[0000-0001-9811-568X]{Adam L. Kraus}
\affil{Department of Astronomy, University of Texas at Austin, Austin, TX 78712, USA}

\author[0000-0002-2580-3614]{Travis A. Berger}
\affiliation{Space Telescope Science Institute, 3700 San Martin Drive, Baltimore, MD 21218, USA}

\author[0000-0001-8832-4488]{Daniel Huber}
\affil{Institute for Astronomy, University of Hawai`i, 2680 Woodlawn Drive, Honolulu, HI 96822, USA}
\affil{Sydney Institute for Astronomy (SIfA), School of Physics, University of Sydney, NSW 2006, Australia}

\correspondingauthor{Kendall Sullivan}
\email{ksulliv4@ucsc.edu}

\begin{abstract}
Identifying and removing binary stars from stellar samples is a crucial but complicated task. Regardless of how carefully a sample is selected, some binaries will remain and complicate interpretation of results, especially via flux contamination of survey photometry. One such sample is the data from the Gaia spacecraft, which is collecting photometry and astrometry of more than $10^{9}$ stars. To quantify the impact of binaries on Gaia photometry, we assembled a sample of known binary stars observed with adaptive optics and with accurately measured parameters, which we used to predict Gaia photometry for each stellar component. We compared the predicted photometry to the actual Gaia photometry for each system, and found that the contamination of Gaia photometry because of multiplicity decreases non-linearly from near-complete contamination ($\rho \leq 0\farcs15$) to no contamination (binary projected separation, or $\rho > 0\farcs3$). We provide an analytic relation to analytically correct photometric bias in a sample of Gaia stars using the binary separation. This correction is necessary because the Gaia PSF photometry extraction does not fully remove the secondary star flux for binaries with separations with $\rho \lesssim 0\farcs3$. We also evaluated the utility of various Gaia quality-of-fit metrics for identifying binary stars and found that RUWE remains the best indicator for unresolved binaries, but multi-peak image fraction probes a separation regime not currently accessible to RUWE.
\end{abstract}


\section{Introduction}\label{sec:intro}
Binary stars are ubiquitous, with about 50\% of main-sequence FGK stars existing in binaries or higher-order multiple systems \citep{Duquennoy1991, Raghavan2010}. Binaries are often viewed as contaminants to surveys, because the addition of a second star to a system can add observational and physical complexities. For example, binaries impact the evolution of circumstellar disks \citep[e.g.,][]{Cieza2009, Kraus2012, Harris2012}, influence the survival and properties of exoplanets \citep[e.g.,][]{Kraus2016, Moe2021, Sullivan2024}, introduce scatter and observational biases into the HR diagram \citep[e.g.,][]{Sullivan2021}, and contaminate magnitude-limited samples \citep{Malmquist1922}.

The over-representation of binaries in magnitude-limited samples is one manifestation of Malmquist bias, which describes how bright objects will be over-represented in magnitude-limited samples because they can be detected out to larger distances. Binaries, which are overluminous for their apparent spectral types, are a textbook example of a population that will exhibit Malmquist bias in its distance-luminosity distribution, because of the additional flux contributed by a secondary star. One example of the impact of Malmquist bias is the complicated stellar selection function of the Kepler Target Catalog (KTC; \citealt{Batalha2010}) for the Kepler space mission \citep{Borucki2010}, where some binaries were excluded and others were over-represented (\citealt{Batalha2010, Wolniewicz2021}, Kraus et al. in prep). 

Binary systems have a wide range of possible separations, and the Malmquist bias for a given system is a function of the binary angular separation: a large binary angular separation ($\rho \gg$ the seeing or diffraction limit) means that the primary will not have any flux contamination from the secondary, while a very small separation (within the seeing limit or diffraction limit of the telescope) will produce total flux contamination. Between the two extrema of complete inclusion to complete exclusion of the secondary flux in photometric measurements of the system, there is likely a gradual decrease in the fractional contribution of the secondary, rather than a bimodal distribution of complete exclusion or inclusion. Quantifying this specific effect in photometric surveys is important to enable corrections of the photometry, especially in large samples. For example, if binaries impact a telescope's flux measurements out to separations of 0\farcs5, the flux correction must consider all binaries with separations out to 0\farcs5, but need not worry about any systems wider than that limit.

One survey where understanding the photometric bias from binaries is particularly important is the Gaia mission \citep{Gaia2016}, now on its third data release \citep[DR3;][]{Gaia2023}. Gaia is a space telescope with a main mission objective of measuring positions and parallaxes for over 10$^{9}$ stars. During these measurements, it also records precise photometry (in one broad filter, the Gaia G band) and spectrophotometry (in two bandpasses, G-RP and G-BP, at R$\sim$100). Precise Gaia photometry, coupled with accurate stellar distances, means that Gaia has quickly become the survey of choice for sample selection in stellar and planetary astrophysics. This makes understanding the systematic behavior of the survey photometry vital for accurately characterizing the data now commonly used in sample selection, including the treatment of stellar multiplicity when extracting photometry.

The Gaia treatment of binary stars varies depending on the system angular separation. \citet{Holl2023} states that equal-brightness binaries with $\rho < 200$ mas are unresolved, $200 < \rho < 400$ mas are ``partially resolved'', and $\rho > 400$ mas are typically resolved. These numbers are bounded by the theoretical Gaia diffraction limit of 150 mas and the nominal effective resolution of 400 mas \citep{Gaia2021}, although completeness of double sources decreases significantly below separations 1\farcs5 \citep{Gaia2021} and with changing binary flux ratios.

Beyond detection of each individual star in the Gaia catalog at sufficiently wide binary separations, some close binaries can also be identified and removed by using cutoffs to the Renormalized Unit Weight Error (RUWE) metric, which is a measure of the excess astrometric noise in the Gaia observations. The literature supports a single-star RUWE cutoff of 1.2-1.4, above which systems are likely binary or otherwise anomalous \citep{Lindegren2018, Belokurov2020, Fitton2022}. At very small separations, some spectroscopic binaries were published in the non-single stars catalog \citep{GaiaBinaries}, and others will be published with complete astrometric solutions in future Gaia data releases. However, each of these detection methods is imperfect, with varying levels of sensitivity to the underlying distribution of binaries \citep[e.g.,][]{Ziegler2018, Penoyre2022}. Thus, even though Gaia metrics present opportunities for identifying binaries in various regions of parameter space, some binaries will inevitably be included in samples selected using Gaia.

Understanding the impact of binaries on Gaia photometry is an important step in correcting for biased photometry when constructing samples with Gaia data. The Gaia spacecraft downlinks windowed ``postage stamp'' images of sources \citep{Rowell2021}, and in DR3 each postage stamp is fit with a single PSF to the source with the global maximum peak flux, with other (local maximum) flux peaks suppressed via masking \citep{Holl2023}. Depending on the precise structure of the mask and the binary separation, some flux from the secondary could still contaminate the primary star PSF, leading to higher measured flux values than should be attributed to the primary star alone, and biasing sample selection and stellar characterization. 

To investigate the possible presence and functional form of the flux contamination from stellar companions in Gaia photometry, we assembled a sample of binary stars with a range of angular separations using the sample of \citet{Sullivan2024}. We retrieved the properties of the individual components of each binary using the methods of \citet{Sullivan2022b} and \citet{Sullivan2023}, then compared the observed Gaia photometry to the predicted single-component photometry for each primary star produced using synthetic photometry to develop a functional form to describe the contamination of Gaia photometry by binary star systems.

\section{The Gaia Mission and Relevant Data Products}
The Gaia mission \citep{Gaia2016} is a space telescope with two 1.45m $\times$ 0.5m apertures, designed for measuring precise astrometry and spectrophotometry for $\sim 10^{9}$ stars. Gaia was launched in 2013 and commenced a 5-year initial mission after commissioning in mid-2014. The final data release (projected to be DR5 in the late 2020s) will include data products from the entire mission, but there have been intermediate releases with progressively more data. The third Gaia Data Release \citep[Gaia DR3;][]{Gaia2023} was made public in 2022, and represents the most advanced Gaia release to date, covering 33 months of the mission duration. Gaia is observing the whole sky in the time domain by scanning across the sky with two perpendicular beams. Crucially for binary star identification, Gaia images stars using $\sim$ 1'' square ``postage stamps'' which are subsequently collapsed along the scan direction before analysis to reduce the impact of smearing from spacecraft motion over the course of the exposure. 

Alongside the basic data products of astrometry and photometry (including colors for most systems), Gaia DR3 included some low- and moderate-resolution spectra, radial velocities, a catalog of non-single stars (mostly comprised of spectroscopic and/or astrometric binaries), solar system objects, and some extragalactic sources. Gaia DR3 also included several metrics for assessing the quality of the data products. These metrics are important because Gaia images themselves are not downlinked, but are instead processed aboard the spacecraft, after which the results are transmitted. Thus, quality metrics are crucial for identifying potential contamination from factors like stellar multiplicity, since neither the individual nor the time series images can be inspected manually.

There are several Gaia metrics that may be able to identify possible stellar multiplicity in different regimes. First, for angular separations outside of $\sim 1''$, most binaries in Gaia are resolved, and the secondary component has a separate entry in the catalog, as long as the secondary star falls above the Gaia brightness cutoff. These binaries can be identified by comparing the 6-D positions of both sources to ensure that they are mutually consistent. For systems with separations $\rho \lesssim 1''$ that are not resolved by Gaia, there are no definitive indicators of stellar multiplicity, although there are several metrics that can identify candidate binaries.

One Gaia metric that can identify candidate binary stars is the Renormalized Unit Weight Error metric, or RUWE \citep{Lindegren2018, Lindegren2021}, which assesses the quality of the astrometric solution. For a well-behaved source, the RUWE value is approximately one. Values of RUWE $\gtrsim$ 1.2 typically indicate a poor astrometric solution, including candidate unresolved binaries \citep{Belokurov2020, Krolikowski2021}, although other factors can also inflate RUWE, such as circumstellar material \citep{Fitton2022}, making follow-up of elevated RUWE systems important for validating candidate binaries. Another potential challenge with using RUWE to identify candidate binaries is that not all systems with a poor astrometric solution have a published RUWE value, making exploration of other possible multiplicity indicators important for assembling a full picture of the impact of stellar multiplicity on Gaia photometry.

Another possible metric for identifying multiplicity from Gaia is the multipeak image fraction, or the \texttt{IPDfmp}. One factor that complicates identification of binary stars in Gaia at binary separations $\rho < 0.5''$ is the unusual shape of the Gaia pixels, combined with the scanning motion of the spacecraft. Gaia pixels are asymmetric, with the rectangle aligned such that the short side of the pixel is parallel to the scanning direction. This choice increases precision and accuracy along the scan. Because the spacecraft orbits, rotates, and precesses as it scans, some fraction of the images of each moderate-separation binary source may include a double-peaked PSF when the postage stamp image is collapsed along the scan direction. Because Gaia images stars multiple times, some fraction of the images will have a multiply-peaked PSF. Thus, \citet{Holl2023} suggest that the \texttt{IPDfmp} is a reliable way to identify some likely binaries in Gaia.

Finally, another parameter that might indicate stellar multiplicity is the photometric flux error itself. The Gaia flux extraction to measure photometry selects a region around the peak flux value in the 1-D postage stamp image with which to extract the flux (i.e., performs aperture photometry), so we hypothesized that some fraction of the measurements may include flux from the secondary stars, while others may not. Because the photometric error is measured as the scatter between measurements in the time series of images, if some images included a component from the secondary while others did not, the error would be inflated as a function of the binary separation and contrast. This impact has not been noted in the literature, so we chose to explore it to see if we could add an additional potential diagnostic to consider while attempting to identify binaries in Gaia data.

\section{Data Collection}
We constructed a sample of 38 binary systems that host Kepler Objects of Interest (KOIs) that we had previously observed and characterized \citep{Sullivan2024}. To briefly recap, to retrieve the individual stellar properties of the components of each binary in our sample, we assembled a data set composed of archival Kepler Input Catalog/2 Micron All Sky Survey (2MASS) photometry \citep[$r'i'z'JHK_{s}$ filters;][]{Skrutskie2006, Brown2011}, flux ratios between the two stellar components (in terms of magnitudes of contrast), and a single unresolved composite stellar spectrum from the Low-Resolution Spectrograph on the Hobby-Eberly Telescope (HET/LRS2) at McDonald Observatory. The following subsections detail the data set used in our analysis.

\subsection{Sample Selection}\label{sec:sample}
We selected our sample using systems from our survey of known binary star exoplanet (candidate) hosts from the Kepler mission \citep[see e.g.,][]{Sullivan2022c, Sullivan2023, Sullivan2024}. The selection criteria for the survey were a planet disposition of ``candidate'' or ``confirmed'', a binary separation $\rho < 2''$ (measured using our high-resolution adaptive optics imaging as described in Sec. \ref{sec:ao}), no higher-order multiplicity with contrast $\Delta {\rm mag} <4.5$ mag within 3.5'', and at least one measured contrast in any filter with $\Delta {\rm mag} < 3.5$ mag. The sample is primarily comprised of binary hosts of small planets. The exoplanet properties are out of the scope of this paper, but were explored in \citet{Sullivan2024}.

To perform the analysis in this work, we chose a subsample of the previously-observed systems restricted to systems with a Gaia contrast $\Delta G < 1$ mag, as calculated using synthetic photometry. This contrast limit was selected to provide the highest quality measurements of the contamination: a contrast of 1 mag corresponds to a flux ratio of $\sim 0.4$, meaning that the secondary flux is a substantial contributor to the total system flux. We tested analyses that included higher-contrast systems, but found that uncertainties from the fainter secondary led to substantial scatter in the final measurements. Similarly, we restricted our sample to only FGK stars (estimated primary star $T_{\rm eff}$ between 4500-6200 K). We ultimately assembled a sample of 38 targets. 

\begin{figure}
    \plotone{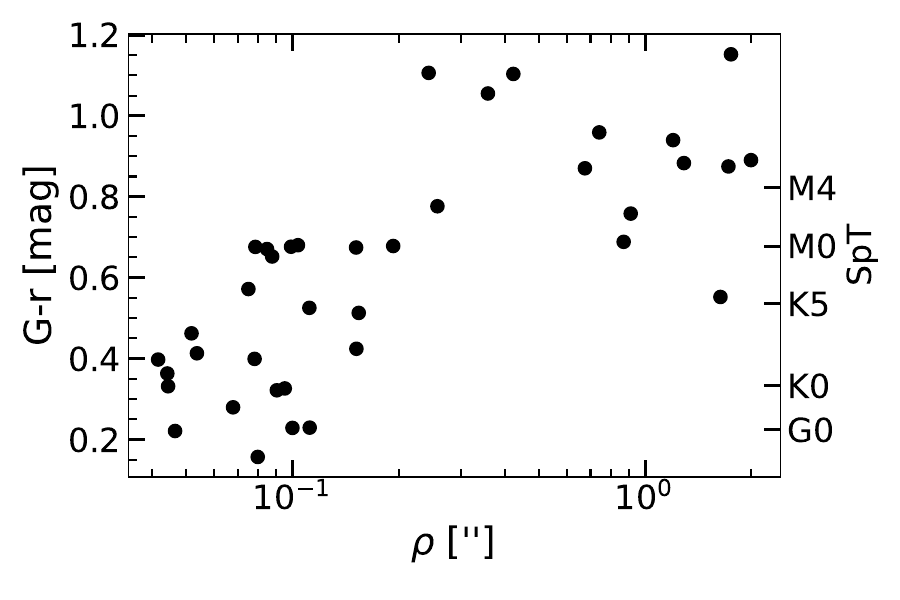}
    \caption{The Gaia G - KIC $r'$ color plotted against binary separation. The right-hand y-axis shows spectral types estimated using synthetic G-r colors from Kraus et al. (in prep), while the left-hand y-axis shows the G-r color for the stars in our sample. At small separations the color is bluer because both Gaia G and KIC r are contaminated by the secondary star, while at wider separations KIC r is contaminated while Gaia G is not, producing a redder color than expected even for mid-M dwarfs.}
    \label{fig:g-minus-r}
\end{figure}

Prior to any additional calculation we plotted the Gaia G - Kepler Input Catalog $r'$ color (KIC; \citealt{Brown2011}) against the angular separation of each source. The $G-r'$ color is expected to be small ($<$0.5 mag) for FGK stars on the main sequence (Kraus et al. in prep). For small separations, where the Gaia G magnitude and the KIC $r'$ magnitude are both contaminated by the presence of the secondary star, the colors are redder than expected for single stars because of the secondary star, but are typically small. However, at wide separations, where the binary is resolved in Gaia but not in the KTC, which had less precise angular resolution, the colors are much redder than expected because of the secondary star, indicating that flux contamination from secondary stars in catalog photometry can result in observed colors that are much redder than expected from single-star relations. This already highlights the impact that correcting photometry for the presence of a secondary star can have on a system's observed magnitude.

\subsection{High-Resolution Imaging Observations}\label{sec:ao}
We only analyzed systems that had existing high-resolution imaging, obtained either by our group or available in the Keck Observatory Archive that were subsequently published in the literature. The data were a combination of speckle interferometry measurements taken in the optical, and adaptive optics (AO) observations taken in the NIR. For the widest systems, there were often resolved measurements from Gaia for each component, in which case we also included the known Gaia G contrast in our analysis of the stellar parameters. 

The speckle measurements were taken from the compilation of \citet{Furlan2017}, which assembled data from several sources, including some observations presented for the first time in that paper. Where available, the sources are cited in Table \ref{tab:obs}. The AO measurements were taken from a combination of the \citet{Furlan2017} compilation and new observations using Keck/NIRC2, presented in \cite{Sullivan2024}. All the contrasts used in our analysis are presented in Table \ref{tab:obs}, along with references.

\subsection{Spectroscopic Observations}
The details of our spectroscopic observations and data reduction are provided in \citet{Sullivan2023}, but we recap briefly here. We observed all systems using LRS2-R, the moderate-resolution (R$\sim1700$) red-optical (6500 $\leq \lambda \leq $ 10500 \AA) arm of the HET facility low-resolution integral field spectrograph \citep{Ramsey1998, Chonis2014, Chonis2016, Hill2021}. The spectra were extracted from the data cubes using an aperture of 2.5 times the seeing value in the wavelength bin with the highest S/N. We corrected the spectra for atmospheric absorption using a least-squares fit to a grid of atmospheric models produced using \texttt{TelFit} \citep{Gullikson2014}. During the parameter retrieval stage, we masked regions of the spectra with significant telluric artifacts, as described in \citet{Sullivan2022c}. The observation dates and S/N of the spectra are listed in Table \ref{tab:obs}.

\section{Parameter Retrieval Techniques}
\subsection{Stellar Parameter Retrieval}\label{sec:stars}
To predict the Gaia magnitude for each component of the binaries in our sample, we retrieved the individual stellar parameters by simultaneously fitting unresolved catalog photometry, high-resolution imaging, and an unresolved spectrum from the HET/LRS2. Our parameter retrieval method using HET spectra is described in \citet{Sullivan2022c}, and the method is explained in detail including validation tests in \citet{Sullivan2022b}. 

Briefly, we simultaneously fit the three component data set of unresolved photometry ($r'i'z'$JHK$_{s}$; from KIC and 2MASS), high-resolution imaging, and an unresolved composite stellar spectrum from HET/LRS2-R. Using the three-component data set we retrieved a set of stellar and system parameters: $\theta = \{T_{1},\ T_{2},\ A_{V},\ R_{1},\ R_{2}/R_{1},\ \varpi\}$, where $T_{1}$ and $T_{2}$ are the primary and secondary star $T_{\rm eff}$, $A_{V}$ is the V-band extinction, $R_{1}$ and $R_{2}/R_{1}$ are the primary star radius and the secondary/primary radius ratio, and $\varpi$ is the parallax. We imposed Gaussian priors on the stellar radii using a 1 Gyr evolutionary model from the MESA Isochrones and Stellar Tracks (MIST) isochrones \citep{Paxton2011, Paxton2013, Paxton2015, Choi2016, Dotter2016}, and a Gaussian prior on the parallax using the Gaia DR3 parallax and error \citep{Gaia2023}. We compared the observed data to the BT-Settl atmospheric models\footnote{\url{https://phoenix.ens-lyon.fr/Grids/BT-Settl/CIFIST2011/}} \citep{Allard2013, Rajpurohit2013, Allard2014, Baraffe2015} with the \citet{Caffau2011} line list.

To fit the model to the data, we initially performed an optimization step using a modified Gibbs sampler that always accepted choices that resulted in a lower $\chi^{2}$ value. We ran the Gibbs sampler until it had reached a total of 200 steps without finding a better fit, then reduced the spread of the distribution used to draw the guesses and continued to draw guesses until another 200 steps had been reached without improvement. Following the optimization stage, we sampled the parameter space using \texttt{emcee} \citep{Foreman-Mackey2013}. We initialized \texttt{emcee} using the best third of walkers from the optimization step, and ran \texttt{emcee} for $10^{4}$ steps or until it converged, using the same convergence criterion as in \citet{Sullivan2023}.

\subsection{Gaia Contrast Calculation}
Using the retrieved stellar parameters, we created synthetic spectra with the appropriate relative flux, temperature, and model-predicted surface gravity for each stellar component. We convolved the predicted synthetic spectra with the broadband Gaia G filter bandpass from (e)DR3, retrieved from the Gaia website\footnote{\url{https://www.cosmos.esa.int/web/gaia/edr3-passbands}}. This convolution resulted in a predicted Gaia G magnitude for each component of each binary, which we could compare to the measured composite system Gaia magnitude. We chose to only investigate the Gaia G flux excess and defer exploration of the BP/RP colors to a future work, because of the additional calibrations needed to convert between our synthetic photometry and the observed values.

\section{Impacts of Stellar Multiplicity on Gaia Measurements}
\begin{deluxetable*}{cCCCCCCCCC}
\setcounter{table}{1}
\tablecaption{Synthetic and Observed Data for Each Source \label{tab:data}}
\tablecolumns{10}
\tablehead{
 \colhead{KOI} & \colhead{$\rho$} & $m_{G}$ & $m_{G,p}$ & $\sigma m_{G,p}$ & $m_{G,s}$ & $\sigma m_{G,s}$ & $\Delta m_{Gaia}$ & $\sigma F_{G}$ & IPDfmp \\
\colhead{} & \colhead{('')} & \colhead{(mag)} & \colhead{(mag)} & \colhead{(mag)} & \colhead{(mag)} & \colhead{(mag)} & \colhead{(mag)} & \colhead{(e s$^{-1}$)} & \colhead{}\\
}
\startdata  
KOI-0163 & 1.20 & 14.11 & 14.06 & 0.04 & 14.44 & 0.05 & 0.379 & 5.86e-09 & 44 \\
KOI-0284 & 0.87 & 12.32 & 12.25 & 0.03 & 12.60 & 0.03 & 0.356 & 3.07e-08 & 67 \\
KOI-0298 & 2.00 & 13.21 & 12.98 & 0.04 & 13.52 & 0.04 & 0.544 & 1.29e-08 & 4 \\
KOI-0349 & 0.15 & 13.75 & 13.92 & 0.06 & 14.40 & 0.08 & 0.478 & 3.40e-08 & 46 \\
KOI-0472 & 0.91 & 15.49 & 15.37 & 0.04 & 15.85 & 0.05 & 0.479 & 1.73e-09 & 63 \\
\enddata
\tablecomments{The full table is available in machine-readable format online.}
\end{deluxetable*}

\subsection{Impacts of Multiplicity on Gaia Photometry}

\begin{figure}
    \plotone{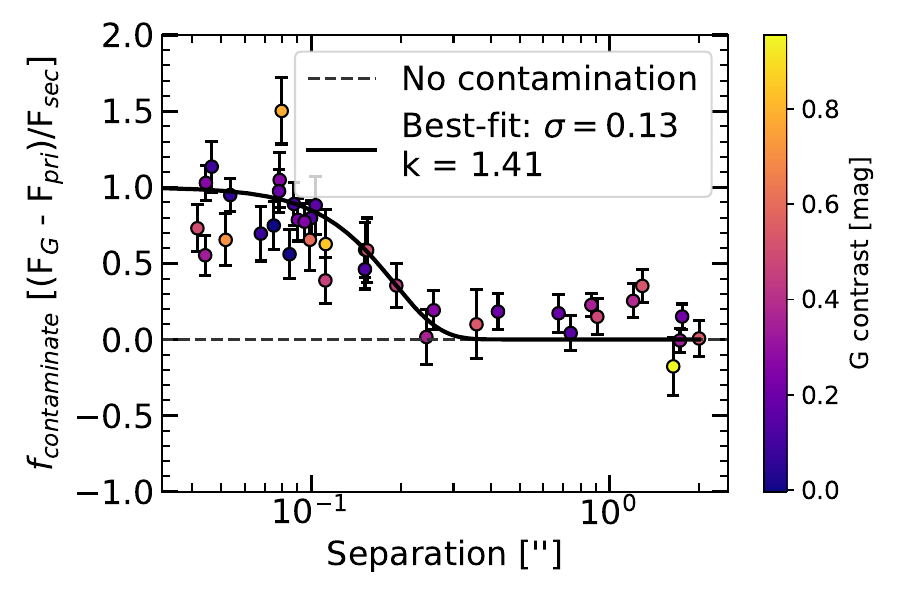}
    \caption{Contamination fraction versus binary angular separation in terms of the secondary star flux. At small separations, nearly all of the secondary star flux is contaminating the Gaia system flux. The secondary star contribution decreases toward zero, reaching approximately zero around 0\farcs4. The zero-point has been adjusted an arbitrary amount to ensure that the resolved binaries at wide separations are centered at zero contamination from the secondary.}
    \label{fig:sep_contamination}
\end{figure}

To understand the degree of contamination (Malmquist bias) from the secondary star for each system, we calculated a contamination fraction, where $f_{\rm contaminate} = (F_{\rm G} - F_{\rm pri})/F_{\rm sec}$. The contamination fraction describes the excess flux in the Gaia magnitude relative to the primary star, in terms of the secondary star flux. For a system with complete contamination from a secondary star, the contamination fraction will be one, while the fraction will be zero for a system with no contamination. We calculated a zero-point for the contamination using a sample of wide systems that definitively fall outside the Gaia binary detection limit ($\rho > 1 \arcsec$, $\Delta G \leq 1$) and adjusted the contamination fraction accordingly such that wide systems had a flux contamination value centered at zero. We calculated the offset as the median of the contamination fraction at wide separations. The offset was $\sim 30$\%, which we attribute to discrepancies between our synthetic photometry (which is produced using the estimated stellar properties as discussed in Section \ref{sec:stars}) and the measured Gaia fluxes resulting from the color bias caused by the presence of a secondary in the KIC, as discussed in Section \ref{sec:sample} and Figure \ref{fig:g-minus-r}.

Figure \ref{fig:sep_contamination} shows the contamination fraction plotted against binary angular separation for our sample. At very small separations the contamination fraction is close to one for all systems, indicating that all the secondary star flux is contaminating the nominal primary star flux measurement, as expected for a fully unresolved system. At wide separations, the contamination fraction is centered at zero by definition, considering our imposed zero-point offset derived from the wide systems and normalized to their median. Between separations of $\sim 0\farcs15$ and $\sim$0\farcs4, the contamination fraction decreases as progressively less of the secondary flux is included in the primary star flux measurement. We fit the contamination fraction with a tunable super-Gaussian of the form 
\begin{equation}
    f_{contaminate} = \exp{\left(-\left[\frac{r^{2}}{2\sigma^{2}}\right]^{k}\right)}.
\end{equation}

The best-fit $\sigma$ and $k$ values are reported in the legend of the plot, and are 0.13$\pm$0.02 and 1.4$\pm$0.6, respectively. The outlier around $\rho$ = 0\farcs1 and contamination fraction $\sim$ 1.5 is an apparently normal small-separation binary, KOI-0635. The additional contaminating flux may indicate the existence of a third star in the system.

\subsection{Binary Detection with Gaia Metrics}
We wished to investigate which, if any, of the possible Gaia diagnostics could effectively identify binary systems. We examined our sample using three Gaia statistics: the RUWE, the multipeak image fraction, or \texttt{IPDfmp}, and the G band photometric variation (parameterized using the error on the G flux). We note that for the majority of these plots the intermediate regime between resolved binaries and systems below the diffraction limit is sparsely sampled, because for many intermediate-separation binaries Gaia DR3 does not report these statistics, limiting their utility as a tool for identifying binaries. However, selecting sources without reported astrometric solutions may be one possible route for identifying intermediate-separation binaries prior to the likely release of full astrometric solutions for those systems in Gaia DR4.

\begin{figure}
    \plotone{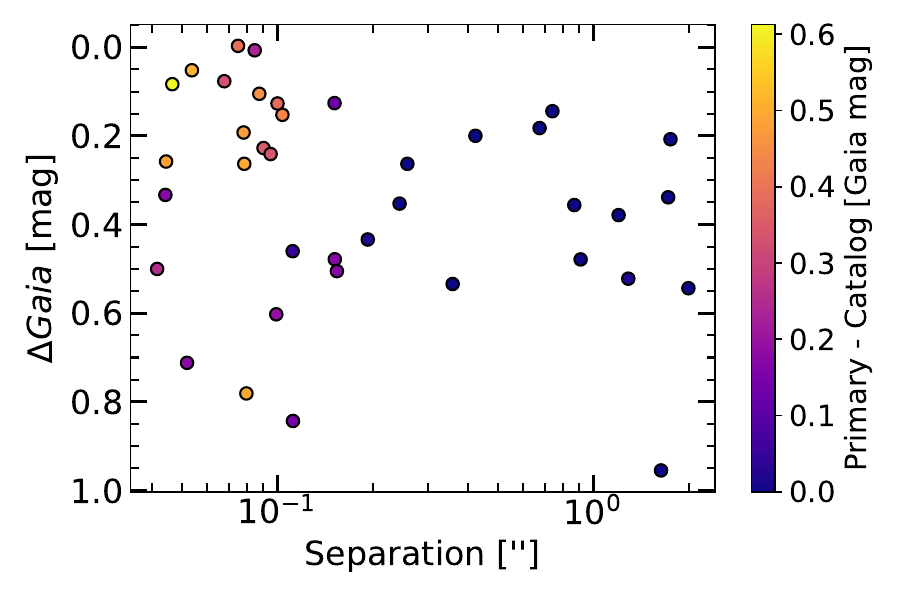}
    \caption{Gaia contrast versus binary separation, with the colorbar showing the residual between the composite and primary star G magnitudes. The residual is correlated with both binary separation and Gaia contrast, verifying that the secondary flux contaminates the system flux and that the impact increases with decreasing separation and increasing contrast, as expected.}
    \label{fig:sep_contrast}
\end{figure}

Figure \ref{fig:sep_contrast} shows a plot of the relationship between binary contrast, binary angular separation, and excess flux. At wide separations and/or large contrasts, there is little-to-no excess flux (presented in the form of a G mag residual between the composite Gaia measurement and the predicted primary star measurement). The residual increases as contrast decreases (i.e., as the secondary gets brighter) and as separation decreases. Thus, it is clear that small-separation and low-contrast systems represent a significant contaminant for Gaia measurements, as quantified in the fit to the contamination fraction shown in Figure \ref{fig:sep_contamination}. 

\begin{figure}
    \plotone{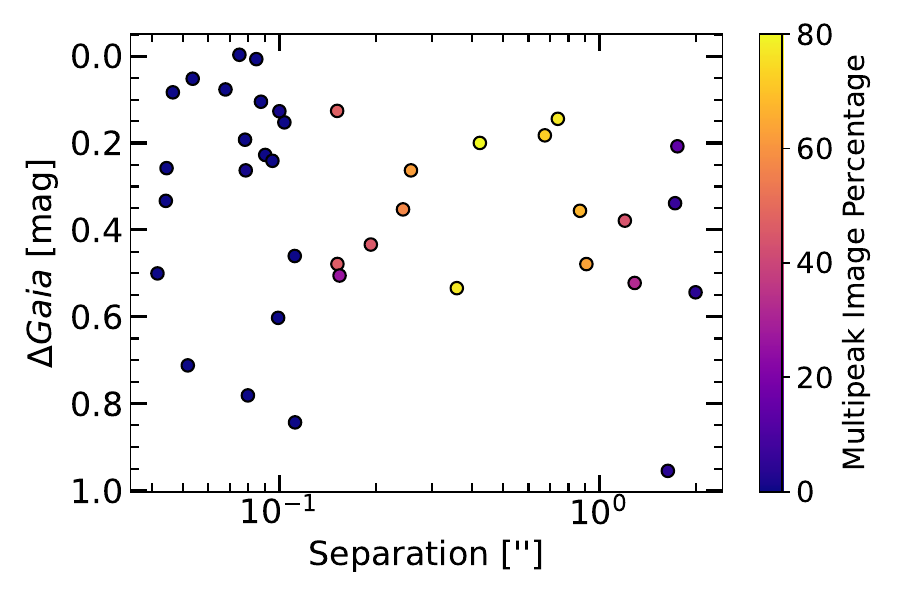}
    \caption{Gaia contrast versus binary separation, with the colorbar showing the Gaia \texttt{ipd\_frac\_multi\_peak} parameter, which tracks the percentage of images showing multiple peaks for a given source. The multipeak percentage increases with decreasing separation, up to a cutoff at $\sim0.''15$, which is the diffraction limit for the Gaia spacecraft. This indicates that for all but the closest binaries, the multipeak percentage is a reliable indicator of unresolved multiplicity.}
    \label{fig:multipeak}
\end{figure}

Figure \ref{fig:multipeak} shows the multipeak percentage (\texttt{ipd\_frac\_multi\_peak} or \texttt{IPDfmp}) for the targets in our sample. The multipeak percentage is the percentage of postage-stamp images of a given system that show multiple flux maxima, and is reported in the Gaia main catalog. The multipeak percentage for our targets increases with decreasing separation down to $\sim 0\farcs15$, which is the diffraction limit of Gaia. This indicates that the multipeak percentage is an excellent indicator of binarity up to the Gaia diffraction limit, as noted in \citet{Holl2023}, and may be a useful counterpart to the RUWE metric, which can also be used to identify unresolved binaries \citep{Lindegren2018, Belokurov2020}.

\begin{figure}
    \plotone{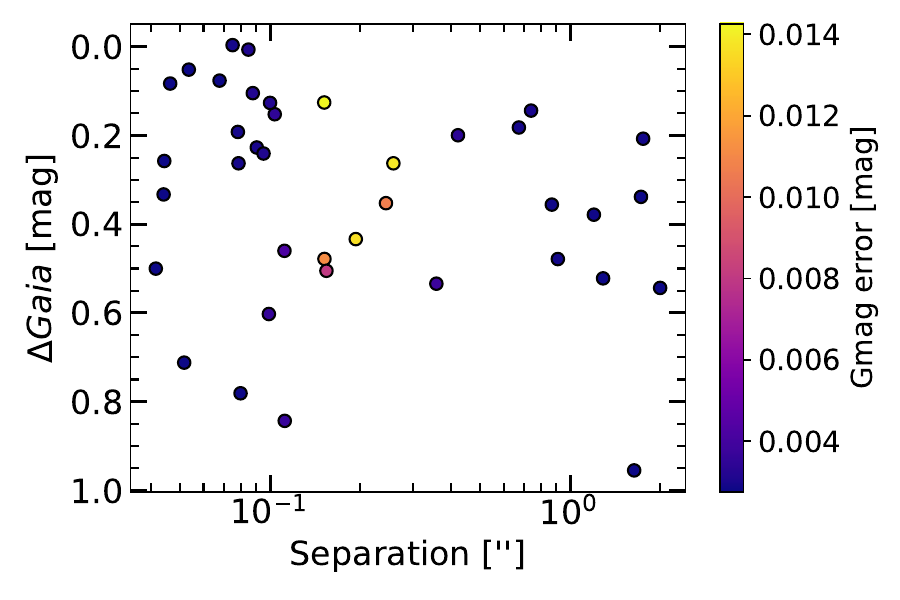}
    \caption{Gaia contrast versus separation, with the colorbar showing the Gaia G magnitude error. The Gaia magnitude error may be correlated with binary separation and contrast, indicating that elevated G mag error may be a weak indicator of stellar multiplicity.}
    \label{fig:gmag}
\end{figure}

We hypothesized that because of the variable scan direction of the Gaia spacecraft, the Gaia G magnitude error may be correlated with multiplicity. This could occur because depending on the scan direction, some fraction of the secondary star flux may be included in the image, altering the apparent G magnitude of the source. Because the reported G magnitude is the average of the multiple photometric measurements, this apparent photometric variability could be reflected in the reported G mag error. We found that the G mag error may demonstrate some sensitivity to multiplicity, but within a much smaller range of separations than the other metrics, such as RUWE and IPDfmp. This lack of sensitivity may be because Gaia observations are collapsed along the along-scan dimension before analysis, reducing how sensitive the photometric error is to multiplicity.

\begin{figure}
    \plotone{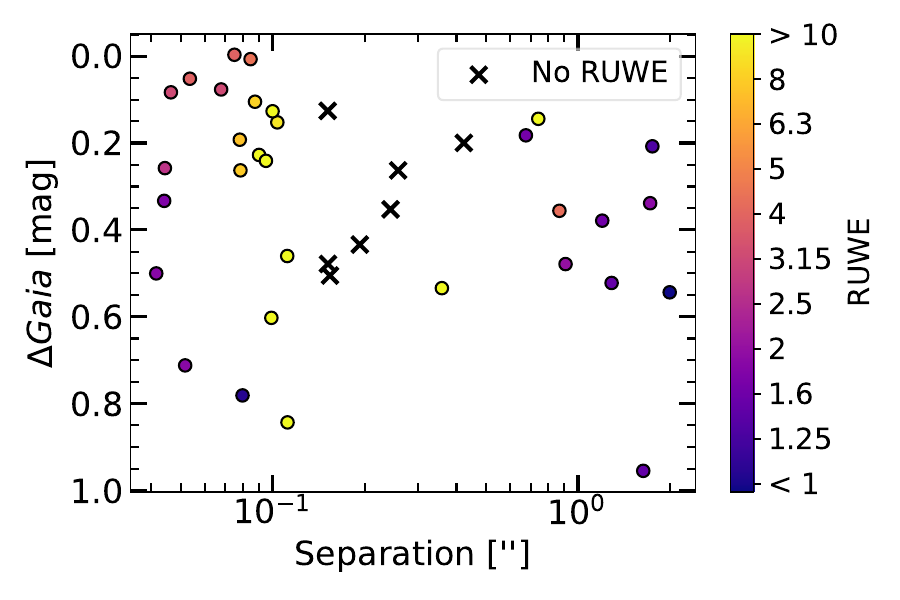}
    \caption{Binary contrast versus separation, color-coded by the Gaia Renormalized Unit Weight Error metric (RUWE). Almost all of the binaries in the sample with separations $\rho < 1''$ have RUWE $>$1.2. However, not all systems have a published RUWE value, especially in the intermediate separation regime, leading to the gap in the separation distribution. Missing systems with no reported RUWE are show in contrast-separation space with black 'x' markers.}
    \label{fig:ruwe}
\end{figure}

Finally, the Gaia Renormalized Unit Weight Error, or RUWE, measures the excess astrometric noise in a system. As with other metrics, elevated RUWE does not definitively identify a binary system, but instead indicates a high likelihood that the system is not typical, which often means multiplicity for field objects. In the literature, an RUWE $>$ 1.2 or 1.4 is typically taken as a cutoff between likely single and likely binary stars \citep{Lindegren2018, Belokurov2020, Penoyre2022}. Within $\rho < 1''$, almost all of the systems in our sample have an elevated RUWE, as demonstrated in Figure \ref{fig:ruwe}. The most substantial elevated RUWEs are systematically between 0\farcs1 and $\sim$0\farcs8, with very close binaries likely not providing enough astrometric variability to significantly inflate the RUWE. Because RUWE is a measure of astrometric noise, RUWE increases as the contrast between the two binary components decreases (i.e., brighter secondaries introduce more astrometric variability). The sparse plot is because not all systems have a published astrometric solution (including a RUWE value) if they are candidate non-single star systems. The regime between the diffraction limit of 0\farcs15 and 0\farcs8 is where some binary systems are marginally resolved \citep{Holl2023}, meaning that many of them do not have published RUWE values.

\section{Discussion and Conclusions}
To explore the magnitude of Malmquist bias and the impact of binary stars in Gaia photometric data, we compared the predicted brightness of the components of binary stars to the measured brightness from Gaia. We found that the flux contamination from secondary stars decreases from 100\% at 0-0\farcs15 to 0\% at 0\farcs3, with a non-linear decline in between. This result indicates that the masking that occurs in Gaia images that have multiple peaks is not complete, and flux contamination from secondary stars still plays a role in the derived photometry.

We found that the degree of contamination is dependent on both the brightness of the secondary star and the binary star angular separation. We also found that the Gaia multipeak image fraction is sensitive to multiplicity, especially at small-to-moderate separations, and that the G magnitude error is only sensitive to stellar multiplicity across a small range of separations. The results regarding RUWE and multipeak image fraction are consistent with the results of \citet{Clark2024}, who investigated the RUWE and multipeak image fraction for a sample of M dwarfs observed using speckle imaging. Our sample of solar-type stars observed at small angular separations with AO is an important additional probe of the smallest-separation binaries.

Our results demonstrate that a variety of metrics provided by the Gaia team can be useful in detecting binaries at different separations. For example, the Gaia multi-peak fraction metric can be used to identify candidate binaries with separations $\sim$0\farcs2-0\farcs5 and moderate contrasts. The RUWE metric appears to be the most effective candidate binary identification method, especially for binaries closer than the the spacecraft diffraction limit of 0\farcs15. However, elevated RUWE can be caused by a variety of effects, and not all Gaia sources have a published RUWE. An elevated RUWE does not fully constrain a hypothetical binary companion, since a variety of mass ratios and separations can produce the same astrometric signal \citep[e.g.,][]{Wood2021}. 

For most systems, RUWE remains the strongest selector for multiplicity, but not all systems have reported RUWE values, and high-contrast systems are less likely to have an elevated RUWE. Therefore, we recommend that a combination of metrics be used to remove small-separation binaries from single-star samples, including a RUWE cut, a spectroscopic binary cut, and cuts using G magnitude error and multi-peak image fraction. These efforts should be supplemented with searches for comoving, cospatial companions at wider binary separations. In cases where removal of binaries is not possible, we recommend using the functional form of flux contamination to impose a flux correction factor by assuming some binary fraction and binary statistics for the sample, then applying the correction to the measured fluxes based on assumed binary population properties. For binaries with known separations in Gaia samples, we recommend using our functional form directly to correct for the flux contamination from the secondary star.

\software{astropy \citep{astropy2013, astropy2018, astropy2022}, matplotlib \citep{Hunter2007}, numpy \citep{Harris2020}, scipy \citep{Virtanen2020}}

\begin{longrotatetable} %
\begin{deluxetable*}{ccCCCCccCCCCCCCc}
\tabletypesize{\tiny}
\setcounter{table}{0}
\tablecaption{System Parameters for Each Source \label{tab:obs}}
\tablecolumns{14}
\tablehead{
\colhead{KOI} & \colhead{$\rho$} & \colhead{HET obs. date} & \colhead{r'} & \colhead{S/N} & \colhead{$\Delta m_{i}$} & \colhead{$\Delta m_{LP600}$} & \colhead{$\Delta m_{Gaia}$} & \colhead{$\Delta m_{562 nm}$} 
& \colhead{$\Delta m_{692 nm}$} & \colhead{$\Delta m_{880 nm}$} & \colhead{$\Delta m_{J}$}  & \colhead{$\Delta m_{K}$} & \colhead{Ref.}\\
\colhead{} & \colhead{('')} & \colhead{} & \colhead{(mag)} & \colhead{} & \colhead{(mag)} & \colhead{(mag)} & \colhead{(mag)} & \colhead{(mag)} & \colhead{(mag)} & \colhead{(mag)} & \colhead{(mag)} & \colhead{(mag)} & \colhead{(mag)} & \colhead{}\\
}
\startdata  
163	& 	1.20	& 	20230910	& 	13.18	& 	324	& 	\nodata & $0.36 \pm 0.03$ & $0.43 \pm $0.05	& \nodata 	& \nodata & 	\nodata & \nodata 	& 	$0.17	\pm 	$0.00	&(1) (4)\\
284	& 	0.87	& 	20210919	& 	11.63	& 	334	& 	\nodata & 	$0.45 \pm 0.04$ & 	$0.36 \pm 0.	00$ & 	$0.00	.81 \pm 	0	15$ & $0.6	7 \pm 0.	19$ & $0.8	5	\pm 0.	$83.00	& $0.2	2	\pm 	$0.03	& $0.26	\pm 	$0.01	&	(2)	(9) (6) (10) (3) (11)\\
298	& 	2.00	& 	20230628	& 	12.32	& 	471	& 	$0.37 \pm 0.01$ & $0.58 \pm 	0.04$ & $0.58 \pm 	0	05$ & \nodata & \nodata & \nodata & $0.22 \pm 0.08$ & $0.19 \pm 0.05$ & (2)	(6) (12) (13)\\
349	& 	0.15	&  20230705 &  13.32 & 349 & \nodata & 	\nodata & \nodata & \nodata & \nodata & \nodata & \nodata & $0.31 \pm 0.05$ & \nodata &(1)	\\
472	& 0.91 & 20230630 & 14.73& 	174	& \nodata & \nodata & $0.58 \pm 	0.05$ & \nodata & 	\nodata & 	\nodata & 	\nodata & $0.27 \pm 0.01$ & (5)	\\
\enddata

\tablecomments{The full table is available in machine-readable format online. The binary separations are from our AO imaging. The references are as follows: (1) = \citet{Sullivan2024}; (2) = \citet{Kraus2016}; (3) = \citet{Furlan2017}; (4) = \citet{Ziegler2017}; (5) = \citet{Ziegler2018}; (6) = \citet{Baranec2016}; (7) = \citet{Law2014}; (8) = \citet{Atkinson2017}; (9) = \citet{Adams2012}; (10) = \citet{Everett2015}; (11) = \citet{Howell2011}; (12) = \citet{LilloBox2012}; (13) = \citet{Wang2015}; (14) = \citet{Dressing2014}.}
\end{deluxetable*}
\end{longrotatetable}

The authors sincerely thank the observing staff and resident astronomers at the Hobby-Eberly Telescope for obtaining some the observations used in this work. The authors acknowledge the Texas Advanced Computing Center (TACC) at The University of Texas at Austin for providing high-performance computing resources that have contributed to the research results reported within this paper.

The Hobby-Eberly Telescope (HET) is a joint project of the University of Texas at Austin, the Pennsylvania State University, Ludwig-Maximilians-Universität München, and Georg-August-Universität Göttingen. The HET is named in honor of its principal benefactors, William P. Hobby and Robert E. Eberly. The Low-Resolution Spectrograph 2 (LRS2) was developed and funded by the University of Texas at Austin McDonald Observatory and Department of Astronomy and by Pennsylvania State University. We thank the Leibniz-Institut für Astrophysik Potsdam (AIP) and the Institut für Astrophysik Göttingen (IAG) for their contributions to the construction of the integral field units. 

Some of the data presented herein were obtained at the W. M. Keck Observatory, which is operated as a scientific partnership among the California Institute of Technology, the University of California and the National Aeronautics and Space Administration. The Observatory was made possible by the generous financial support of the W. M. Keck Foundation. The authors wish to recognize and acknowledge the very significant cultural role and reverence that the summit of Maunakea has always had within the indigenous Hawaiian community. We are most fortunate to have the opportunity to conduct observations from this mountain. 

This publication makes use of data products from the Two Micron All Sky Survey, which is a joint project of the University of Massachusetts and the Infrared Processing and Analysis Center/California Institute of Technology, funded by the National Aeronautics and Space Administration and the National Science Foundation. This research has made use of the SVO Filter Profile Service (\url{http://svo2.cab.inta-csic.es/theory/fps/}) supported from the Spanish MINECO through grant AYA2017-84089. This research has made use of the VizieR catalogue access tool, CDS, Strasbourg, France (DOI : 10.26093/cds/vizier). The original description of the VizieR service was published in 2000, A\&AS 143, 23. This work has made use of data from the European Space Agency (ESA) mission {\it Gaia} (\url{https://www.cosmos.esa.int/gaia}), processed by the {\it Gaia} Data Processing and Analysis Consortium (DPAC, \url{https://www.cosmos.esa.int/web/gaia/dpac/consortium}). Funding for the DPAC has been provided by national institutions, in particular the institutions participating in the {\it Gaia} Multilateral Agreement.
D.H. acknowledges support from the Alfred P. Sloan Foundation, the National Aeronautics and Space Administration (80NSSC22K0781), and the Australian Research Council (FT200100871).

\bibliography{gaia_bib}
\end{document}